# A SPATIAL DATA MODEL FOR MOVING OBJECT DATABASES

Hadi Hajari and Farshad Hakimpour

Department of Geomatics Engineering, University of Tehran, Tehran, Iran

## ABSTRACT

*Moving Object Databases will have significant role in Geospatial Information Systems as they allow users to model continuous movements of entities in the databases and perform spatio-temporal analysis. For representing and querying moving objects, an algebra with a comprehensive framework of User Defined Types together with a set of functions on those types is needed. Moreover, concerning real world applications, moving objects move along constrained environments like transportation networks so that an extra algebra for modeling networks is demanded, too. These algebras can be inserted in any data model if their designs are based on available standards such as Open Geospatial Consortium that provides a common model for existing DBMS's. In this paper, we focus on extending a spatial data model for constrained moving objects. Static and moving geometries in our model are based on Open Geospatial Consortium standards. We also extend Structured Query Language for retrieving, querying, and manipulating spatio-temporal data related to moving objects as a simple and expressive query language. Finally as a proof-of-concept, we implement a generator to generate data for moving objects constrained by a transportation network. Such a generator primarily aims at traffic planning applications.*

## 1. INTRODUCTION

Nowadays hardware such as position-aware devices or on-board units in vehicles is being used extensively and prompts many new kinds of geospatial applications. These applications produce large amount of movement information that should be managed and analyzed optimally in database systems and be made available for spatio-temporal analysis. Current database systems are far from handling such tasks and new techniques in spatial databases are needed.

Attempts to add such ability to databases have created a new domain in spatio-temporal databases entitled Moving Object Databases (MODs). Research on the field of MODs has begun since late 1990s, and is recently receiving a lot of interests due to the prominent developments in positioning technologies. These databases deal with the geometries that change over time continuously, also called spatio-temporal data [1]. In this context two important abstractions are moving points and moving regions in which only time-dependent locations, or also time-dependent shapes or extents need to be managed, respectively. Examples of moving points and moving regions are GPS-equipped cars, people or ships and hurricanes, forest fires or the spread of an illness.

The main concern of this paper is the moving geometries in which shapes or extents are invariable during movements. On the other hand, regarding significant applications in a number of scientific domains, we consider moving objects constrained by a transportation network instead of free movement. Some typical moving object application fields based on transportation networks specifically in Geospatial Information Systems (GISs) are travellers' trip planning, traffic management, and Location-Based Services (LBS).

Since commercial databases have widespread usage in various types of applications, designing and developing a database purely for moving objects applications imposes extra cost. The efficient way is to design algebras for moving objects applications and embed them into existing DBMS data models. This objective will be fulfilled if the algebras' designs are in accordance with common standards between existing DBMS data models such as Open Geospatial Consortium (OGC) [2]. The OpenGIS Simple Feature Access, also called ISO 19125, describes a common architecture to store and manage simple feature geometries. In this context, there is also another standard called ISO/IEC 13249 SQL/MM [3]. This standard standardizes extensions for multi-media and application-specific packages in Structured Query Language (SQL). Part 3 of the standard is the international standard that defines how to store, retrieve and process spatial data using SQL. Comparing these standards is beyond the scope of this paper. For more information read [4]. We integrated moving object concept within OGC specification because it has been adopted by the most popular DBMSs' implementations such as ORACLE Spatial, PostGIS, MYSQL, and DB2 Spatial.

In this paper, we present OGC-based models consisting of relations for modeling spatial networks and a number of User Defined Types (UDTs) and operations for optimal storage, representation, and querying of static and moving features on the extensibility interface of Object-Relational database systems.

Additionally, we implement the developed models on top of the Oracle database given that it is a widely-used and pioneer DBMS supporting OGC-compliant spatial standard. This implementation demonstrates the capabilities of the models not only for modeling transportation networks but also for representing and querying moving objects. At last, we propose a prototype for simulating cars' motions on street networks to evaluate the presented data model.

The rest of this paper is organized as follows. We start with some notes on general MOD traits and then present works related to our contribution in the next section. Afterwards, we present the network data model and respective UDTs and operations for representing static points and lines on the network. In section four, we describe UDTs and operations for modeling and querying moving points on the network. We discuss implementation aspects of the model in Oracle 11g in section five. Section six assesses the applicability of the proposed system via introducing a data generator. We discuss the results of the system and illustrating a few example queries in section seven. Finally, the paper ends with conclusion remarks in section eight.

## 2. RELATED WORKS

In this section, we first discuss the most strictly related work on modeling and querying moving objects in spatial networks. Then the data generators which provide data sets for moving objects on transportation networks are introduced. At the end, we give a comprehensive and thorough view on other research areas in the field of MODs.

Two perspectives of moving object databases in the past years are location management and spatio-temporal database [5]. The first one focuses on answering questions on the current positions of moving objects, and on their predicted temporal evolution in the near future. In this perspective, the Moving Objects Spatio-Temporal (MOST) model and the Future Temporal Logic (FTL) language have been proposed in [6][7]. The models in the second perspective deal with the trajectories or the complete histories of moving objects. This view was pursued, for example, in [8][9]. In this view the complete evolution of a moving object can be represented as an attribute within the databases. In this article we concentrate on the second perspective as we intend to show complete history (past and current) of moving points' movements in OGC-based ORDBMSs. Also, concerning moving points' applications in real world, we intend to represent movement information on spatial networks.

One of the major standard models for networks is Geographic Data Files (GDF). In [10], this standard is explained comprehensively. In a nutshell, GDF is a European standard used to describe road networks and road-related data. According to this standard, a road network comprises a set of Roads which are defined in terms of Road Elements that they contain. A Road Element is the smallest unit of the road network and having a Junction at each end and a Junction is located at the intersection of two or more road centrelines. All in all, road network model in GDF has a three-level structure:

Level 0: Topology. At this level everything has been described by nodes and edges.
Level 1: Features. Level 1 describes all the simple features as Road Elements and Junctions. Features can have attributes that are specific (i.e. one way, road width, number of lanes).
Level 2: Complex Features. At this level the "simple features" are aggregated to a higher-level feature. Level 2 is mostly used when a straightforward description of the road network is enough, e.g., give guidance instructions to the driver.

In this context most noticeable investigations are as follows. Grumbach and his colleagues developed a spatial database system called DEDALE which relied on a constraint-based model to represent and manipulate geometric and spatio-temporal data [11]. An important opinion that has not been considered in this research is that in many cases objects do not move freely in the 2D space but rather within spatially embedded networks. A paper by [12] considers modeling and querying moving objects in road networks. The network is modelled using usual graph models and consists of a relation of blocks (edges). A generic network model is not defined formally here. Moreover, some data modeling issues for spatial networks have been looked in [13][14][15]. Jensen and his colleagues explain that real road networks are quite complex and a simple directed graph is not sufficient. They used several interrelated representations for modeling networks. Although a well-formed spatial network is introduced here, it's too complicated to be a good basis for a query language.

In 2006, Güting provided a data model and query language for objects moving in networks [8]. The data model has some interesting features and we have drawn some of our motivation from this paper. The most noticeable parts of this model summarized here.

In this data model, positions of moving points are described relative to the network instead of the embedding space, i.e., the positions are represented within road networks using Linear Referencing System (LRS) rather than x, y of the points, so the geometry of the points is stored once with the network. For instance, finding out connection between moving objects and parts of the network can be done much simpler by checking the route identifier in the description of the movement. Otherwise complicated and time-consuming operations must be done. Moreover, users can extend the information about the network by using the standard facilities of the database systems, e.g. users can add or remove some information on the network relations about speed limits on some parts of the network or restricting turning to right or left in a junction. Different kinds of routes (e.g. highways, one-way, or two-way routes) can be defined in this data model and special operations for network such that shortest path can be executed. Besides, the model allows users to describe static or moving objects relative to the network, such as motels, gas stations, vehicles, traffic jam, etc.

We take advantage of the elements mentioned above and altered the proposed data model according to common standards between DBMS data models. The developed network data model in this paper conforms to GDF standard. The modified moving object's data model balance storage efficiency and expressive power also provide adequate support for the algorithms that process the data. Moreover, the data model gives this ability to the users to add data to the existing information.

The data model introduced in [8] is implemented in the SECONDO DBMS. It's still not feasible to use this database in commercial applications because of its incompatibilities with SQL. So we are motivated to make some alterations to the data model according to OGC standards. The modified data model can be embedded into database systems easily. Adding such a data model to existing DBMSs allows users to not only do spatio-temporal analysis about moving objects movements but also use the traditional facilities of the databases.

Toward this direction, Pelekis and his colleagues have proposed HERMES system by two papers in [16][17]. Likewise, another Oracle extension, named STOC (Spatio-Temporal Object Cartridge), is presented in [18] to support spatio-temporal data management in Oracle. These systems are implemented on top of Oracle spatial data model based on a set of basic data types, together with spatial data types, temporal data types (TAU Temporal Literal Library), and some UDTs for moving entities. There are still critical differences between our contribution and these prototypes. Firstly, there is no formal representation and modeling of spatial networks. As we mentioned before, we purpose to model spatial networks explicitly in accordance with OGC standards, therefore the model can be inserted into the kernel of DBMS data models or as an extra package, e.g. ORACLE, POSTGRES, DB2, and MYSQL. Secondly, we consider distances on transportation networks with their complexities (e.g. one-way links, U-turns, limiting in turning to right or left at a junction, speed limit on parts of a network, etc.).

Now we introduce some works on existing data generators. Data generators are used in providing data sets for moving objects. Many data generators are developed for moving entities in recent years such as GSTD [19], OPORTO [20], and SUMO [21]. But in this part we present the most popular generators which provide data for moving objects constrained by a transportation network. An open-source generator with a visual interface for objects moving along transportation networks is proposed in [22]. During the generation process, speed and route of a car depend on the load of network edges.

Christian Düntgen and his co-workers have utilized SECONDO DBMS, instead of developing dedicated generator software, to generate moving object data on transportation networks [23]. The data model implemented in SECONDO system is based on free movement in the 2D plane. However, the above mentioned data generator provides network constrained data. The implementation on SECONDO data model is in progress for supporting objects moving on network.

In this article, we aim to provide a data generator for evaluating the proposed data model. We develop a user-friendly generator that provides data for cars moving on spatial networks. The raw behaviour of the generator is controlled by a set of parameters. In this generator cars are produced at random (following a creation rate function) and terminate as soon as they reach their destination.

We now pass on to other fascinating research areas in the field of moving object database both from a geoinformatics as well as database research perspective. In the early 2000s, the field of MODs has blossomed and a lot of research in various areas has been done especially on uncertainty in moving objects trajectories, indexing of spatial trajectories, mining on trajectory data and Moving Objects Ontology (MOO). In the following, we address some studies in each area.

The sources of uncertainty in the moving object trajectories are abundant, for example, the imprecision of the positioning devices and interpolation between consecutive location sampling. Therefore, many researchers have focused on addressing the problem of uncertainty management in moving object databases. Some works on modeling of uncertainty together with managing and querying uncertain data in MODs are [24][25][26].

Alongside the modeling and representing uncertainty, it is momentous that MODs can support the retrieval of trajectory data effectively and efficiently. To extract traveling history of moving objects in MODs containing very large numbers of trajectories, the efficiency depends crucially upon an appropriate index of trajectories. Hence a lot of papers have been done on implementation issues and developing index structures [27][28][29][30].

Another research area in the field of MODs is trajectory mining. Normally, moving objects trajectories are available as sample points (x,y,t), which have very little semantics. The analysis and knowledge discovery from trajectory sample points is very difficult from the user's point of view, and there is an emerging need for new data models and tools to extract meaningful patterns from these data. Substantial applications in diverse domains need to identify and utilize groups of trajectories that exhibit similar patterns from a collection of trajectories. Example applications include transportation optimization, prediction-enabled services, scientific and social analysis applications, sports analyses, as well as crowd and outlier analyses [31]. In [32], a semantic trajectory data mining query language for extracting meaningful patterns from trajectories has been proposed. There has also been work developing several algorithms for mining trajectory sample points and discovering similar trajectories [33][34][35]. In addition, some papers [36][37][38] studied similarity search for moving object trajectories in spatial networks.

Besides, semantic trajectory is an appealing research area that has recently emerged in GIS to enhance the modeling and analysis of moving object trajectory [39][40][41]. A semantic approach for pattern discovery in trajectories is proposed in [42]. This approach that relies on ontology enhances object movement information with event semantics. Moreover, in [43] event-based approaches are presented to capture the dynamic aspects of moving objects with descriptions of the semantics of real-world occurrences. Some other studies in the research domain of MOO are [44][45]. Although lately many data models and algorithms are developed in the last two areas, mining moving object and MOO, there are still deficiencies in many aspects, e.g. on trajectory clustering in spatial networks and use of ontology to infer knowledge.

# 3. NETWORK DATA MODEL AND OPERATIONS

We design the spatial network data model as a set of junctions, routes, and sections. This data model has the ability to be manipulated through some operations by the users. Furthermore, we introduce user defined data types based on the network data model for displaying static points and lines on the network and useful operations on these data types that allow users to interact with the network.

## 3.1. Spatial Network Modeling

Here a transportation network is associated with a set of sections, routes and junctions between routes. SECTION class, the most basic element in designing the network data model, stores the graph structure of the transportation networks. The graph structure should be available to support network-specific operations. Also other constitutive classes of the network (routes and junctions) are formed based on the section class. Section class is comprised of an identifier, the route identifier that the section forms part of it, start and end nodes' ids of the section, measures of the start and end position of the section on the route, kind of the section (one-way or two-way), length and LRS geometry of the section (see UML class Section in Figure 1). Values will be stored in instances of this class at two stages: after importing raw data, e.g., from ESRI Shapefiles and after creating routes.

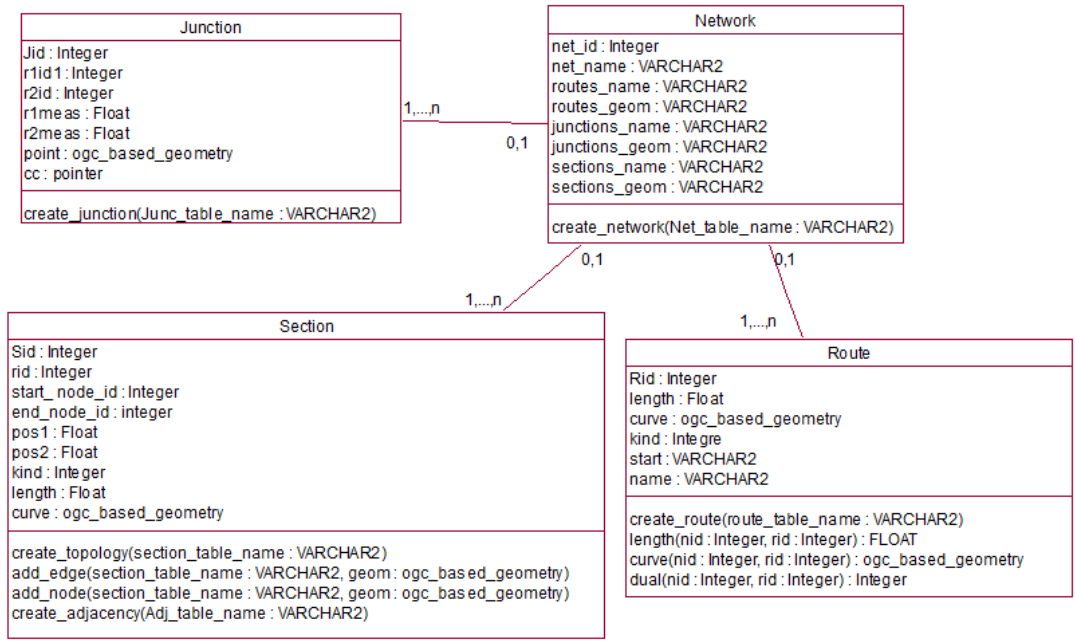

Figure 1.  UML class diagram of the network data model

The routes can be bi-directional, i.e., adopt movement in two directions and it may be necessary to distinguish positions on the both sides of a route. As described before, we use LRS for capturing data object locations within road networks, i.e., the positions of moving objects on roads are just distances from the origin of the routes, instead of, using x, y of the objects. As a result, we store geometries of the moving objects just once for all with the network. A route description consists of an identifier, a length, a route type (1 for an one_way route and 2 for a two_way route), a flag indicating how route locations have to be embedded into space (from start or end of the route), and a curve describing its geometry in the plane that is represented by an OGC-compliant geometry. Figure 1 shows associations between Network, Junction, Section, and Route classes and the respective operations for each class.

A junction in a network consists of two route measures with distinct route identifiers and a connectivity code (cc), an attribute encoding which movements through the junction are possible (See Junction class in Figure 1). Obviously, if the junction is consisted of three different routes then three distinct junction instances of this class will be created. The geometry of the junction in the plane is stored in point attribute in the class. Algorithms of the operations which are based on user-defined constraints such as computing shortest path on the network benefit from connectivity code at each junction. For each route the *up* direction of movement is distinguished where the LRS measure of the route is increasing and *down* direction is when the measure is decreasing. Figure 2 (a) shows a junction in which the transitions $B_{up} \rightarrow A_{down}$, $B_{up} \rightarrow A_{up}$, $B_{down} \rightarrow A_{down}$, $B_{up} \rightarrow B_{down}$ (U-Turn) are allowed and Figure 2 (b) represents the possible transitions in a 4×4 matrix.

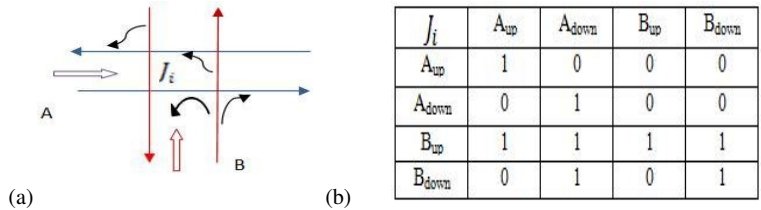

| $J_i$ | A_{up} | A_{down} | B_{up} | B_{down} |
|---|---|---|---|---|
| A_{up} | 1 | 0 | 0 | 0 |
| A_{down} | 0 | 1 | 0 | 0 |
| B_{up} | 1 | 1 | 1 | 1 |
| B_{down} | 0 | 1 | 0 | 1 |

Figure 2. (a) Diagrammatical representation of a junction, (b) its transition matrix

## 3.2. Data types on network

In this subsection, we introduce user-defined data types, GPOINT and GLINE, for representing static positions and regions within the network (e.g., motels, gas stations, speed limit, under construction area and etc.) based on the concepts described in the network data model in previous subsection.

A GPOINT contains a position on a route possibly on one of its sides. The UML class diagram for defining a GPOINT object and the operations for accessing its values is represented in Figure 3.

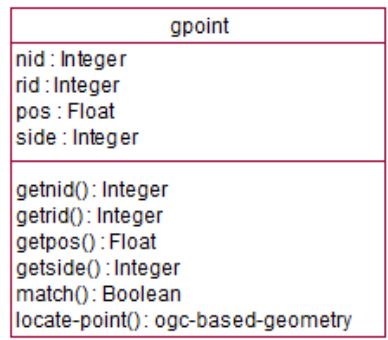

Figure 3. UML class definition of a static point on the network

A position on the network consists of an identifier for the network (if there are some networks in the database system), the route identifier on which the point is located, the distance from the origin of the route (measure), and a side value that is represented by integers: 1 for the right side of the route's up direction and -1 for the left side of the route's up direction. LOCATE-POINT operation returns OGC-based geometry of the position with specified measure on the network.

A GLINE object describes a region of the network, for example, the part of the network within a fog or speed limit area, or the output of the shortest path algorithm. Therefore it may consist of some route intervals in the network in which they are composed of some pairs of route locations on the same or different routes. If a GLINE value consists of some route intervals, they have to be quasi-disjoint. Two route intervals are quasi-disjoint if they are either on different routes, or on the same route with different sets of route's locations. The UML class diagram for defining a GLINE value and the operations for accessing its values is represented in Figure 4. A GLINE value like a GPOINT contains a network and route identifier, pos1 and pos2 show the start and end positions of the constitutive route intervals of the GLINE on the route, and glid keeps the GLINE identifier. If a GLINE is composed of some route intervals, all of them will be stored in the relation with the same glid. This attribute is useful when analysts want to retrieve all parts of a network with specific characteristic for example, finding speed limit parts on the network.

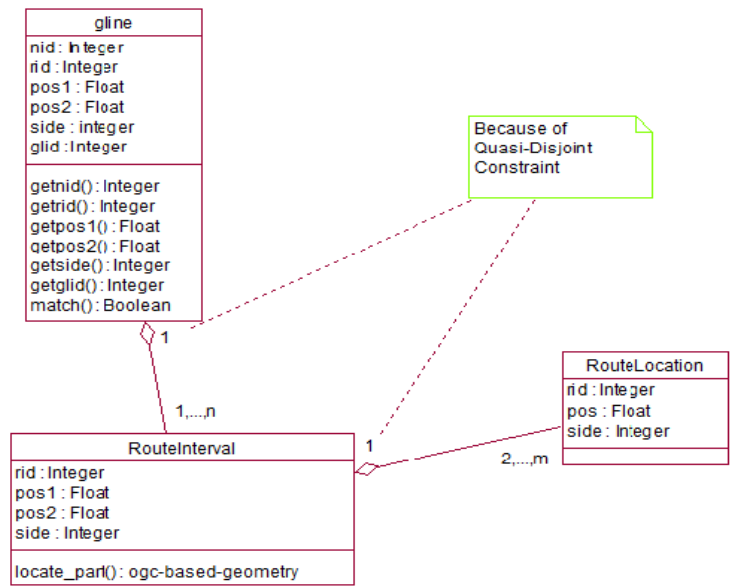

Figure 4.  UML class diagram of a static line with quasi-disjoint constraint

## 3.3. Operations on network

In this subsection, first we present essential operations for creating and updating the graph structure of the networks. Moreover, we define operations that allow users to get more information about the network and the objects on it. Two major categories of operations in this part are accessing routes' information on the network and checking topological relationships between data types and network. In the end, we describe the most important network-specific algorithm SHORTEST_PATH.

CREATE_TOPOLOGY operation calculates the connectivity relationships between nodes and edges of the imported data automatically. In this operation, we use existing methods at DBMS's spatial data model to find spatial interaction between geometries.

Before using CREATE_TOPOLOGY operation, the line shapefile should be imported to the database at the section_table_name table. This operation adds two attributes named start_node_id and end_node_id to the table, then calculates and stores values of connectivity relationships in start_node_id, end_node_id and Sid attributes. Two other operations namely ADD_EDGE and ADD_NODE add an edge and a node to the topology respectively and the existing connectivity relationships will be updated automatically.

CREATE_TOPOLOGY (Varchar2 section_table_name)
ADD_EDGE (Varchar2 section_table_name, OGC-based-geometry edge)
ADD_NODE (Varchar2 section_table_name, OGC-based-geometry node)

Other defined operations on network are implemented based on the attributes of the user-defined types (GPOINT, GLINE) and return the ideal outputs. Users can access routes information simply by the following operations. LENGTH operation uses the length method in the spatial data model and returns the length of a route with specified route_id. CURVE operation returns the LRS geometry of the route and DUAL operation returns kind of the route.

LENGTH (int network_id, int route_id)   → Float
CURVE (int network_id, int route_id)   → OGC-based-geometry
DUAL (int network_id,  int routeid_id)   → int (kind of the route)

Moreover, some operations for checking the topological relationships are defined below. A GPOINT can belong to a route. A GLINE value can intersect, be contained in, or contains a route.

ON_ROUTE (gpoint geom, int route_id)   → Boolean
INTERSECTS (gline geom, int route_id)   → Boolean
CONTAINS (gline geom, int route_id)   → Boolean
IS_CONTAINED (int route_id, geom gline)  → Boolean

For computing the shortest path between two points on the network, JUNCTIONS, ROUTES, and SECTIONS relations are needed. Also we have to gain access to a persistent adjacency list data structure. The purpose of the adjacency list data structure is to find outgoing edges from a node. Adjacency list structure is automatically produced from connectivity code attribute at JUNCTIONS class. We adapt Dijkstra's algorithm considering the user-defined constraints on spatial network (e.g., no u-turn at a junction). While Dijkstra's explore all possible paths, A* tries to look for a better path by using a heuristic function which gives priority to nodes that are supposed to be better than others. Since A* optimality depends on the heuristic function used, we found Dijkstra's algorithm is more appropriate to be modified for computing path with minimum cost and considering spatial network constraints. In Figure 5 the data structure for the junction of Figure 2 and the adjacency list structure are represented.

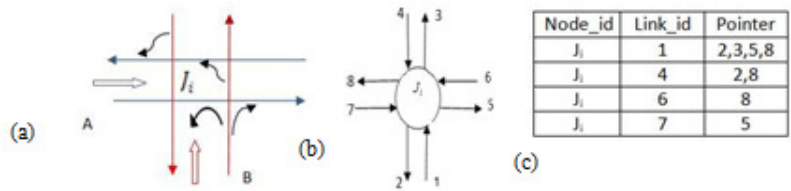

Figure 5.  Adjacency list structure for a junction; (a) a junction, (b) the node and incident links,
(c) the adjacency list data structure

If the data model in the database supports network modeling and network constraints can be applied to the network data model, existing shortest path function in the data model by accepting two node ids as their input arguments computes the distance between the nodes. However, the counterpart operation in this paper has the ability to compute paths with minimum costs between two locations (not necessarily two nodes) on the network.

In summary, the network model introduced above conforms to GDF standard described at section 2. The proposed network model consists of a set of junctions and routes which they contain sections (Road Elements). Operations such as CREATE_TOPOLOGY, ADD_EDGE, AND ADD_NODE, which are used for creating and modifying network topology, are implemented based on the network data structure at Level 0 defined in GDF standard. That is, network is constructed of individual nodes and edges. Furthermore network data structure at Level 1 is used to calculate shortest path between two locations along the transportation network. At this level network constraints are applied and stored in adjacency list data structure. Finally other developed operations on network utilize the network data structure at Level 2 because these operations don't require high level of details about the transportation network configuration.

# 4. DATA TYPES AND OPERATIONS FOR MODELING MOVING ENTITIES ON NETWORK

In this section we present data type MGPOINT for modeling moving geometries on the transportation network and auxiliary time data types. Also operations on these data types will be introduced through them users can query and analyze spatio-temporal data in a simple and efficient way. The operations can be classified in the following categories: projection operations that restricts moving points' movements to spatial and time domain like AT, AT_INSTANT, ATPERIODS, DEFTIME, TRAJECTORY; topological operations concerning topological relationships between static and moving types like INSIDE, INTERSECT; direction and distance operations like SHORTEST_PATH, DIRECTION; and numeric operations that compute a numeric value like SIZE, NOW. Besides, operations are defined for interaction between network space and space like IN_SPACE and IN_NETWORK.

## 4.1. Moving point geometry data type

We use a new user-defined data type called MGPOINT for modeling moving points, e.g. vehicles, on the network. For this UDT we need to store the network and route identification, a time interval, two positions (measures) on the network showing the positions at the beginning and the end of the time interval, the side of the route, and the ID of the moving object to recognize each object individually. Storing values about velocity and acceleration of moving objects at each period is critical, in particular, where positions of objects between times $t_1$ and $t_2$ should be interpolated or when trajectory mining methods are used to evaluate average traffic rate over particular periods of time at specified areas.

To be able to implement data type MGPOINT, we apply a discrete model, i.e., in this data type we use sliced representation to show continuous movements of entities. Each slice is associated with a temporal unit and describes the movements of the object during the time interval via measures on the network. Defining finite representations of moving data types (moving point, moving lines, moving regions) has been introduced in [46]. Figure 6 shows UML class definition of this data type with respective operations.

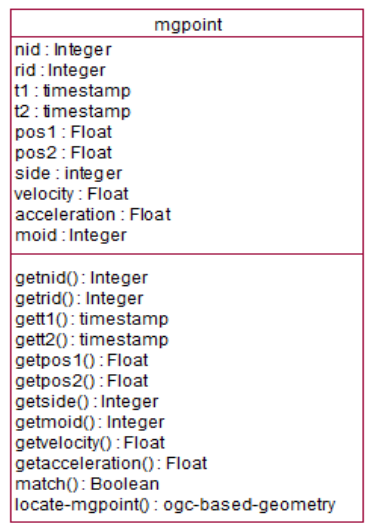

Figure 6. UML class definition for a moving point with operations

To finalize the data model for moving objects on top of an OGC-based ORDBMS, we present some additional data types in Table 1 including available timestamp data type in database.

Table 1.  Auxiliary data types

| |
|---|
| PERIOD (timestamp × timestamp) |
| INTIME (gpoint × timestamp) |
| PERIODS is a collection of time intervals |
| UGPOINT is a collection of mgpoint |

PERIOD data type is used for bounding two time intervals, for example, PERIOD ('01.01.2011 21:24:42:600','dd.mm.yyyy hh24:mi:ss:ff3'), ('01.01.2011 21:30:03:009','dd.mm.yyyy hh24:mi:ss:ff3') shows the time is bounded between two specified time in a day. INTIME data type is used for representing the position of a moving car in a specified time. In the next subsection, we describe operations which are defined on these UDTs.

## 4.2. Operations on moving point geometries

In this subsection we provide the bulk of operations on the defined data types in our model. In Table 2, different categories of operations which are applied to our defined data types are shown. Here we describe the semantics of the operations concisely. IN_SPACE and IN_NETWORK operations are defined for converting between network and spatial values. INST and VAL operations return the two components of an Intime value. DEFTIME operation returns all time intervals during which the object is defined. TRAJECTORY operation yields the traversed part of the network by a moving object. AT_INSTANT operation returns the position of the moving object at a specified time. AT_PERIODS returns part of the network travelled by a moving object in a time interval. AT operation restricts a function to the times when its value lies within the second argument. DIRECTION operation returns direction of movement at a particular time. INSIDE operation checks whether a moving point is located within part of the network at the specified time or not. SIZE operation calculates total length of a trajectory. It is useful when traffic analysts aim to consider how many kilometres each cab is travelling during a day. DURATION operation computes time span when the moving object begins and ends its movement. CURRENT and NOW operations return the last position and the last stored time of the moving object that is moving on the network, respectively.

Table 2.  Spatial, temporal and spatio-temporal operations

IN_SPACE (gpoint or gline geom) → OGC-based-geometry

IN_NETWORK (OGC-based-geometry geom) → gpoint

VAL (intime position) → gpoint

INST (intime position) → timestamp

DEFTIME (Ugpoint positions) → periods

TRAJECTORY (Ugpoint positions) → gline

ATINSTANT (int moid, timestamp time) → intime

ATPERIODS (int moid, periods time) → Ugpoint

DIRECTION (int moid1, int moid2, timestamp time) → Float

SHORTEST_PATH (int moid1, int moid2, timestamp time) → gline

AT (int moid, gline geom) → Ugpoint

INSIDE (int moid, gline geom, timestamp time) → Boolean

SIZE (gline geom) → Float

DURATION (Ugpoint geom) → Float
NOW (int moid) → timestamp
CURRENT (int moid) → gpoint

In this part, the main extensions to the Güting's algebra [8] include:

Representing, storing, and retrieving moving object's locations are extended by OGC geometries. Operations in this paper work directly with OGC-based geometries: e.g., IN_NETWORK operation allows users to input OGC-based geometries instead of measure, IN_SPACE (GPOINT geom) operation produces OGC-based geometry based on measure value on the network, and overloaded IN_SPACE (GLINE geom) operation returns spatial OGC-based geometry of a path. In the end, some modifications in the signature of operations are developed to make them more comprehensible for the end users and permit them to manage and query trajectories more easily.

In Figure 7, we describe an extension of the OGC hierarchy that integrates new user-defined data types which are devoted to managing continuous movement of moving objects. Operations for GPOINT, GLINE, ROUTEINTERVAL and MGPOINT classes are provided in their UML class diagrams (Fig 3, 4, 6).

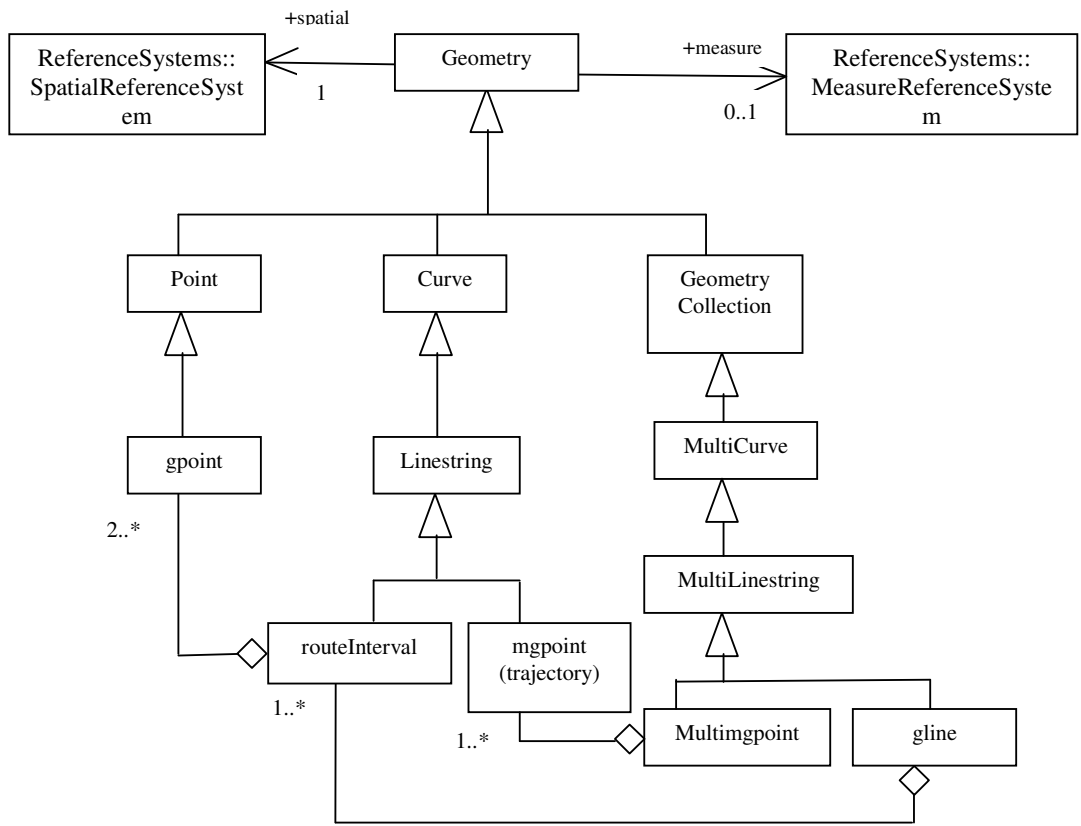

Figure 7. Extended OGC class hierarchy for moving objects

Figure 8 shows overview of our model described in previous sections. Our moving object engine uses standard data types existing in any databases' data models such as Varchar, Timestamp, Number, and OGC-based spatial geometry. In addition, the engine makes use of our defined data types comprising of network and auxiliary data types together with operations acting on top of these data types.

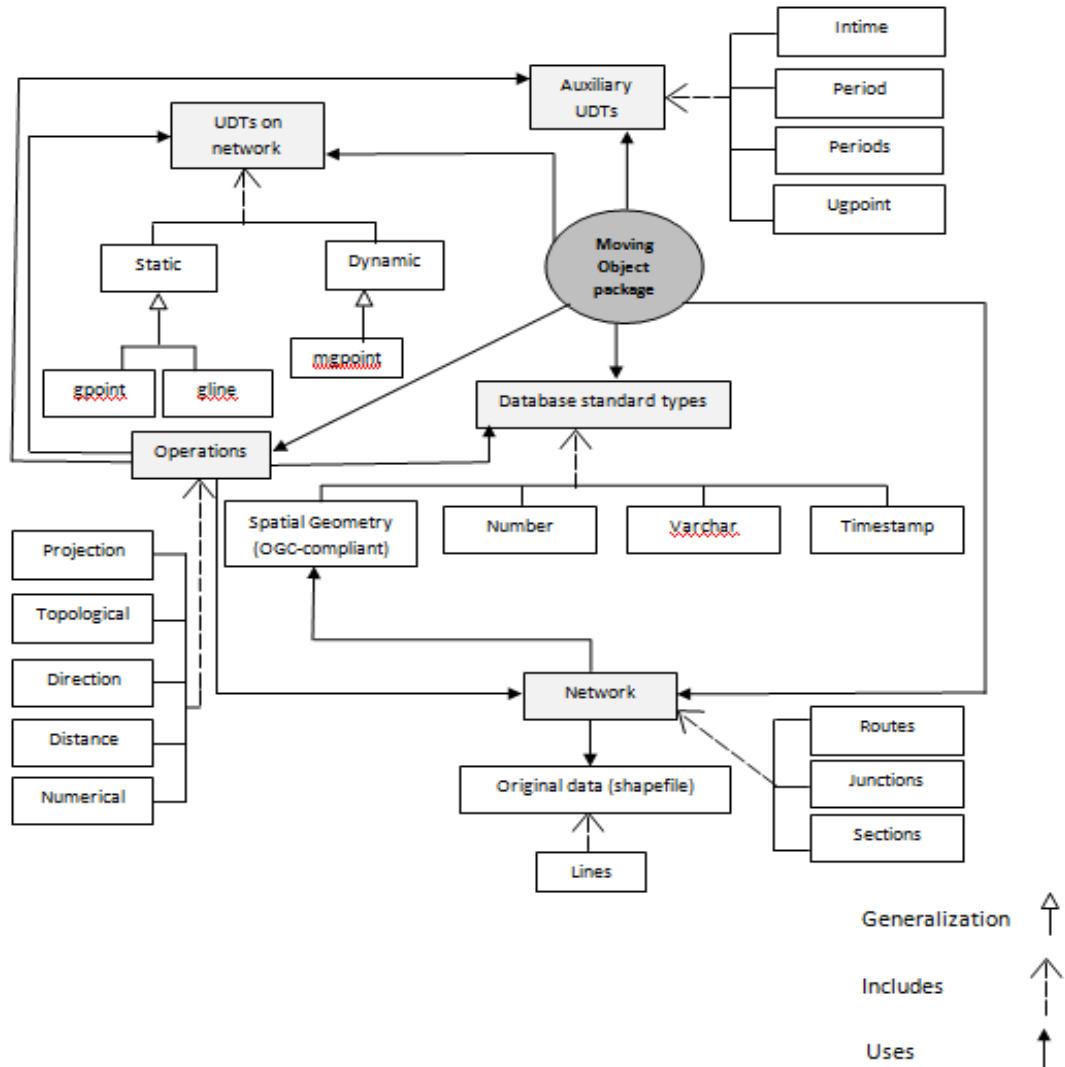

Figure 8.  Overview of the MO Engine

# 5. IMPLEMENTATION

In this section we address some implementation issues of the proposed framework on top of the OGC-compliant Oracle spatial data model. Oracle data model consists of two general types of networks: OGC-compliant spatial and non spatial network data models [47]. A non spatial network data model includes connectivity information but no geometric information such as a logical network. A spatial network contains both connectivity information and geometric information such as SDO network and LRS network. In a SDO network, the nodes and links contain SDO_GEOMETRY geometry objects without measure information or with measure information in a LRS network. Transportation road networks are good instances of the second type of the Oracle network data model. Oracle spatial network data model has the capability to represent network restrictions (e.g., prohibited turns in a road network) and therefore apply them on network analysis.

We extend Oracle data model with adding a MO engine consisting of building the network data model, creating UDTs, and operations via Java programming language and PL/SQL. In the MO engine, we use standard data types in Oracle, OGC-compliant Oracle spatial geometry, temporal and spatio-temporal UDTs on network, and operations that are developed on the data types. To increase performance of the package, some of these operations invoke developed JAVA functions. As a result, we provide a package that allows users to build the network, query and analyze movements of entities.

## 5.1. Building network and describing implementation aspects

In this subsection, first we describe how to implement spatial network data model in the database system. Secondly we present data structures for the data types GPOINT, GLINE, and MGPOINT. We built the network data model of a small district through ROUTES, JUNCTIONS, and SECTIONS tables. As described before, values in table SECTIONS will be caught at two stages. When the original spatial data that consists of edges is imported to the database, we build topology from the spatial geometries with CREATE_TOPOLOGY in the developed MO package. Then table ROUTES will be generated using CREATE_ROUTE operation via the developed java extensions. Afterwards, we can store other related information about each network section in table SECTIONS that are beginning and end measures of each section on the associated route. Finally we form junctions and adjacency list data structure with the respective operations CREATE_ADJACENCY and CREATE_JUNCTION in the package. Constructing theses tables according to standard data types in Oracle is shown in Table 3.

Table 3. Constitutive tables of the network

| Create table tbl_sections (sid number, rid number, $pos_1$ number, $pos_2$ number, kind number, length number, curve sdo_geometry, constraint cnt_sec_pk primary key(sid)) |
| --- |
| Create table tbl_routes (rid number, kind number, length number, curve sdo_geometry, start varchar2(8), name varchar2(100), constraint cnt_rou_pk primary key(rid)) |
| Create table tbl_junctions (jid number, $r_1$id number, $r_2$id number, $r_1$meas number, $r_2$meas number, point sdo_geometry, cc nested table, constraint cnt_junc_pk primary key(jid), nested table cc store as junc_cc) |

Now we are ready to represent static and moving entities along the transportation network based on the network data model. Users can store all of static objects like motels, gas stations, and malls at the NETWORK_GPOINT table which consists of a column GEOM for collecting stationary point geometries. Also users keep data about static linear entities on the network such as foggy or speed limit parts at NETWORK_GLINE table. The GLINE type constructor is slightly more complex, since it contains a finite set of quasi-disjoint route intervals. We put a trigger on this table to check the quasi-disjoint constraint when new records are being inserted or some old records are being updated. For representing data type MGPOINT, we follow the UML class definition in Figure 6. We save the positions of the moving objects on the network in the table NETWORK_MGPOINT.

Table 4 shows SQL codes for creating NETWORK_GPOINT, NETWORK_GLINE, and NETWORK_MGPOINT. Some pseudo codes for creating GPOINT, GLINE, and MGPOINT types at the MO package are shown in Table 5, which are associated with some member functions.

Table 4. Forming tables for keeping static and moving entities

| Create table tbl_network_gpoint (id number, geom gpoint, name varchar2(100), constraint cnt_gpoint_pk primary key(id)) |
|---|
| Create table tbl_network_gline (id number, geom gline, name varchar2(100), constraint cnt_gline_pk primary key(id)) |
| Create table tbl_network_mgpoint (id number, geom mgpoint, constraint cnt_mgpoint_pk primary key(id)) |

Table 5. Pseudo codes for operations trajectory and atinstant

| Create or replace function Trajectory (Integer moid) Return gline Find all of tuples at network_mgpoint table that store movements of the object with specified moid Extract pos1,pos2,side,routeid,netid attributes From the selected tuples Specify a unique glid for the output Save the values at the gline type |
|---|
| Create or replace function atinstant (Integer moid, timestamp tm) Return intime Search in network_,gpoint table and find all of tuples having the specified moid Check moid had been moving at the specified time If true Find the period that includes the tm time Interpolate the position of the moid at the tm time according to the velocity and acceleration of the object at that period Save the position and other related information of the intime type |

Additionally, with regard to strong arguments mentioned in section 2 about representing locations along a spatial network with measures rather than x and y, Oracle LRS network data model can be a suitable alternative for proposed network data model in this paper. Of course, it is needed to modify comprising LRS network tables by adding some attributes to match with the UML classes we introduced in Figure 1. As a result, data types and operations can be developed based on this network model.

Toward this direction, we altered the OGC-based Oracle network data model and tested the developed operations and data types according to the new network data model. In short, we created an LRS geometry network in Oracle and altered the network tables with adding extra data columns to conform to our network framework. In addition to using existing operations in our data model, we make good use of other network analysis in Oracle network data model including shortest path, nearest neighbour, within cost, travelling sales man problem and so on. Here network analysis computations can be restricted by network constraints. For example, during the shortest path analysis, we implemented turn restrictions at certain junctions. As a result, we illustrated that the developed algebra for operations has the flexibility to work with any OGC-based network data model. If a database data model, for example postgres, cannot model transportation networks, the first developed algebra in this paper can be easily utilized for network modeling.

## 6. DATA GENERATOR

To evaluate our work, we develop a data generator that simulates trajectories of moving objects on transportation networks. Such a benchmark is useful in several ways: it provides well-defined data sets and queries for experimental evaluations; it points out weaknesses of the data model or implementation issues of the algorithms and motivation for further research; give a chance to the users how to analyze spatio-temporal data through the user defined data types and functions. To generate the data set, we simulate a number of cars driving on the road network for a given period of time and capture their positions every two seconds. Finally generated data sets will be stored at the table NETWORK_MGPOINT in the database system.

Figure 9 shows the flowchart of the simulation process. A trip is parameterized by a triple (Start, End, Start_time), where Start and End are the origin and the destination nodes respectively and are chosen using a uniform distribution among all nodes of the network, and Start_time is the instant when the trip begins. Trips will be created along the network considering its constraints (e.g., turn-right or turn-left prohibition or U-turn at an intersection) according to the developed shortest path operation.

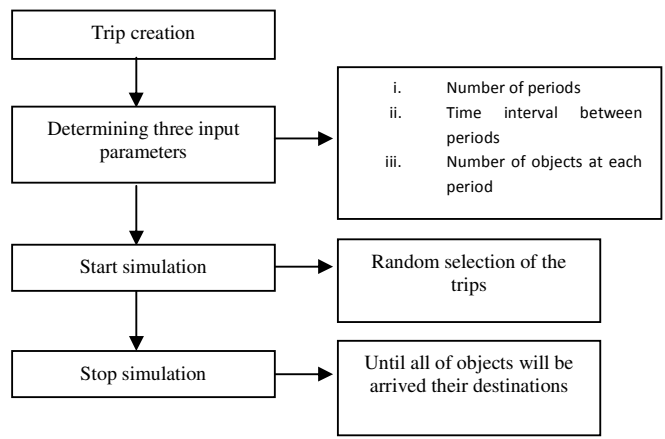

Figure 9. Flowchart of the simulation process

The simulation process begins with three parameters: number of periods for generating objects, time interval between periods, and number of objects that will be generated at each period. During the simulation process, objects move with a constant velocity (Equation 1). They slow down when approaching a junction or stop if they are required to do so by a red traffic light or objects accelerate when leaving a junction (Equation 2). This process continues until all of objects arrive at their destinations.

$$x = 0.5 \times a \times t^2 + v_0 \times t + x_0 \qquad\qquad (1)$$

$$v = a \times t + v_0 \qquad\qquad (2)$$

One of the most important issues in mobility scenarios is updating the moving objects' positions that change continuously. Keeping position data up-to-date in database systems is a must. On the other hand, frequent updating would also impose serious storage and performance problems for the database. We consider this matter in the data generator system, i.e., the moving point geometries will be recorded in the database when objects are entered new routes or objects' accelerations are changed. Positions of the moving entities between time instants stored in the database can be determined by interpolation (using equ1). Figure 10 illustrates designed interface of the generator.

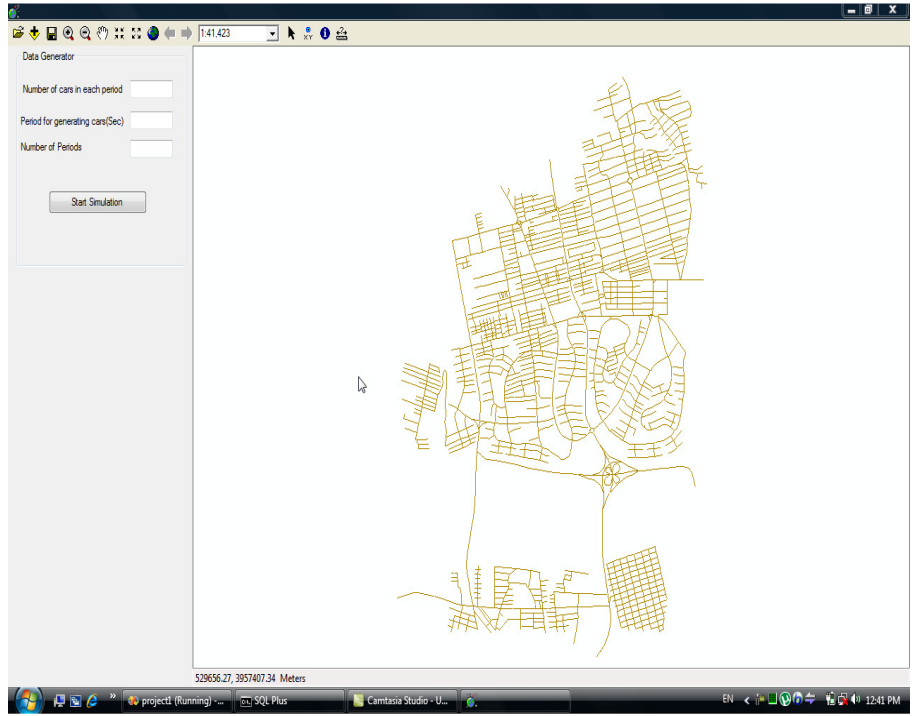

Figure 10.  Data generator's interface

## 7. RESULTS

In this section we show a set of queries for the traffic management application. The ESRI line shapefile used in this prototype includes 35000 links. The road network of the line shapefile is created at the database according to the constitutive tables of the network data model. Also we have tables network_gpoint describing static point entities and network_gline describing regions of the network. We simulated movements of 100 cars in 20 periods with time interval 25 seconds. The number of cars created at each period is 5. Table network_mgpoint keeps movement data of cars that are being created by the generator. The main goal of this part is to give an idea about how to formulate queries in moving object applications in SQL using the provided MO package. We try to pose queries about complete history (past and current) of cars' movements.

**Query 1:** Which places did the car with license plate 1033 visit between 00:04:42:600 and 00:10:03 of last Saturday?

DEFINE lastsaturday= "periods (to_timestamp ('21.01.2011  00:04:42:600' , 'dd.mm.yyyy   hh24:mi:ss:ff3') , to_timestamp ('21.01.2011 00:10:03:000' , 'dd.mm.yyyy hh24:mi:ss:ff3'))";

SELECT trajectory (atperiods (&lastsaturday, 1033)) FROM network_mgpoint c
WHERE c.geom.moid = 1033;
**Output:**                     Trajectory_GL(GLINE(1,19,635.031,2.952,-1,1033),GLINE(1,19,2.952,0,-1,1033),
GLINE(1,131,969.939,989,0630,1,1033),GLINE(1,131,989.630,1755.836,1,1033),
GLINE(1,131,1755.836,1760.889,1,1033),GLINE(1,143,0,17.041,1,1033), GLINE(1,143,17.041,374.177,1,1033))

**Query 2:** Find all cars that visited Chamran highway between 23:50 and 00:11 of yesterday?

DEFINE yesterday="periods  (to_timestamp  ('27.01.2011  23:50:00:000'  ,  'dd.mm.yyyy    hh24:mi:ss:ff3')  , to_timestamp ('27.01.2011 00:11:00:000' , 'dd.mm.yyyy hh24:mi:ss:ff3'))";

SELECT c.moid FROM network_mgpoint c WHERE
Intersection (Chamran, trajectory (at (Chamran,  atperiods (c.geom.moid, &yesterday)))) = 'TRUE';

**Output**: 1000, 1005, 1011, 1031, 1036, 1055, 1061

**Query 3:** Which street has more than a hundred cars currently?

SELECT c.routeid, c.name
FROM routes c, network_mgpoint v
WHERE on_route (c.routeid, current (v.geom.moid)) = 'TRUE'
GROUP BY c.routeid
HAVING COUNT(*) > 100;

**Output**:

| routeid | name |
|---------|----------|
| 143 | Valiasr |
| 198 | Karimkhan |
| 298 | Kargar |

# 8. CONCLUSION AND FUTURE WORKS

The main contribution of this paper is developing a precise and comprehensive data model and query language for moving objects in networks on top of an OGC-compliant DBMS data models. In more detail, in this paper we provided a comprehensive framework for modeling transportation network considering real world network complexities. Also, we developed UDTs for presenting moving objects on the network. Finally, to allow users to query and analyze spatio-temporal data we developed suitable operations on these UDTs. In section 5, we illustrated that the developed algebra for operations has the flexibility to work with any OGC-based network data model. That is, if a DBMS's data model supports spatial network modeling, there is no more need to use our proposed network model. Consequently, the developed algebra for the set of data types and operations can be implemented and worked on the DBMS's built-in network model efficiently. As a proof of concept, we modified tables in LRS geometry network data model in Oracle and tested the operations and data types which are defined in sections 4 and 3 according to the new network data model. As another result we have extended SQL-like query languages for querying moving objects.

Besides, for evaluating the query language and operations, we simulated a system of cars' motions. We have presented a method to generate a data set representing cars driving on transportation networks. The number of cars and the observation times can be changed easily. This evaluation demonstrates that the system extension provides a simple, powerful, and expressive language for querying spatio-temporal data. Moreover, this simulation can be a good case study for traffic management in GIS as a real application. Future work will address implementation issues such as improving algorithms for the operations. Also we would like to extract patterns from moving objects' movements by integrating semantic geographic information and trajectory information. Moreover, we are attempting to propose an OGC standard that encodes properties describing motion of an object moving through city streets, such as velocity, acceleration, heading, timestamp, and position, on the Internet. The GML encoding can be extended in this candidate standard.